\documentclass[twocolumn,english,conference]{IEEEtran}
\usepackage[T1]{fontenc}
\usepackage[latin9]{inputenc}
\usepackage{url}
\usepackage{amsmath}
\usepackage{amssymb}
\usepackage{babel}

\makeatletter

\newcommand{\QQ}{\mathbb{Q}}
\newcommand{\RR}{\mathbb{R}}
\newcommand{\ZZ}{\mathbb{Z}}
\newcommand{\CC}{\mathbb{C}}

\newcommand{\cv}{\mathbf{c}}
\newcommand{\hv}{\mathbf{h}}

\newcommand{\rv}{\mathbf{r}}
\newcommand{\sv}{\mathbf{s}}

\newcommand{\uv}{\mathbf{u}}
\newcommand{\vv}{\mathbf{v}}
\newcommand{\wv}{\mathbf{w}}
\newcommand{\xv}{\mathbf{x}}
\newcommand{\yv}{\mathbf{y}}
\newcommand{\zv}{\mathbf{z}}

\newcommand{\Oc}{\mathcal{O}}

\newcommand{\Uc}{\mathcal{U}}
\newcommand{\Vc}{\mathcal{V}}

\newcommand{\vol}{{\rm Vol}}

\newcommand{\diag}{\mathsf{diag}}
\newcommand{\snr}{\mathsf{SNR}}
\newcommand{\vect}{\mathsf{vec}}

\makeatother

\usepackage{babel}

\begin{document}

\title{Lattice Code Design for the Rayleigh Fading Wiretap Channel}

\author{\authorblockN{Jean-Claude Belfiore} \authorblockA{Department
of Communications and Electronics \\
 TELECOM ParisTech\\
 Paris, France\\
 \textbf{Email:} belfiore@telecom-paristech.fr} \and \authorblockN{Frédérique
Oggier} \authorblockA{Division of Mathematical Sciences \\
 School of Physical and Mathematical Sciences\\
 Nanyang Technological University, Singapore\\
 \textbf{Email:} frederique@ntu.edu.sg} }
\maketitle
\begin{abstract}
It has been shown recently that coding for the Gaussian Wiretap Channel can be done with nested lattices. A fine lattice intended to the legitimate user must be designed as a usual lattice code for the Gaussian Channel, while a coarse lattice is added to introduce confusion at the eavesdropper, whose theta series must be minimized. We present a design criterion for both the fine and coarse lattice to obtain wiretap lattice codes for the Rayleigh fading Wiretap Channel. 
\end{abstract}

\section{Introduction}

\subsection{Related work}

The wiretap channel was introduced by Wyner \cite{Wyner-I} as a
discrete memoryless broadcast channel where the sender, Alice, transmits
confidential messages to a legal receiver Bob, in the presence of
an eavesdropper Eve. Wyner defined the perfect secrecy capacity as
the maximum amount of information that Alice can send to Bob while
insuring that Eve gets a negligible amount of information. He also
described a generic coset coding strategy, where both data and random bits 
are encoded, in order to confuse the eavesdropper (see also \cite{Wyner-II}).
The question of determining the secrecy capacity of many classes of
channels has been addressed extensively recently, yielding a plethora
of information theoretical results on secrecy capacity (see 
\cite{Shlomo} for a survey of many such results).

There is a sharp contrast with the situation of wiretap code designs,
where very little is known. The most exploited approach to get practical
codes so far has been to use LDPC codes (see \cite{TDCMM-07}
for binary erasure and symmetric channels, \cite{KHMBK-09} for Gaussian
channels with binary inputs). 
Finally, lattice codes for Gaussian channels have been considered
from an information theoretical point of view in \cite{HY-09}.

A design criterion for constructing explicit lattice codes on the Gaussian Wiretap channel has been proposed in \cite{sec-gain},
based on the analysis of Eve's correct decision probability. This
design criterion relies on a new lattice invariant called {}``secrecy
gain'' which is based on the lattice theta series. The secrecy gain of unimodular 
lattice was further studied in \cite{unimodular}.  

\subsection{Contribution and organization}

We propose here to find the appropriate design criterion for both the wiretap fast fading and block fading channels and to give some intuition on lattice codes which are optimal for this criterion.

This paper is organized as follows. Section II presents the system model and recalls 
the design criterion for the Gaussian wiretap channel. Sections III and IV are the main contributions where we give the code design criterion for, respectively, the fast fading and the block fading channel. The particular case of algebraic lattices is discussed in both cases. 

%
%

\section{System Model and the Gaussian case}

\subsection{Fast fading channels}

Alice wants to send data to Bob on a wiretap fading channel, where
an eavesdropper Eve is trying to intercept the data through another
fading channel. Perfect channel state information (CSI) is assumed at both 
receivers. Thus it is possible to remove the phase of the complex fading coefficients to obtain a real fading which is Rayleigh distributed, with the aid of an in-phase/quadrature component interleaver to guarantee that the fading coefficients are 
independent from one real symbol to the next \cite[sec~2.1]{Oggier-2}.
This is modeled by 
\begin{equation}\label{eq:fastfading}
\begin{array}{ccl}
\yv & = & \diag(\hv_{b})\xv+\vv_{b}\\
\zv & = & \diag(\hv_{e})\xv+\vv_{e},
\end{array}
\end{equation}
where $\xv\in\RR^n$ is the transmitted signal, $\vv_{b}$ and $\vv_{e}$
denote the Gaussian noise at Bob, respectively Eve's side, both with
zero mean, and respective variance $\sigma_{b}^{2}$ and $\sigma_{e}^{2}$,
and 
\begin{equation}\label{eq:fadingmatrix}
\begin{split}\diag(\hv_{b})=\left(\begin{array}{ccc}
|h_{b,1}|\\
 & \ddots\\
 &  & |h_{b,n}|\end{array}\right),~\\
\diag(\hv_{e})=\left(\begin{array}{ccc}
|h_{e,1}|\\
 & \ddots\\
 &  & |h_{e,n}|\end{array}\right)\end{split}
\end{equation}
are the channel matrices containing the fading coefficients where
$h_{b,i},h_{e,i}$ are complex Gaussian random variables with variance
$\sigma_{h,b}^{2}$, resp. $\sigma_{h,e}^{2}$, so that $|h_{b,i}|,|h_{e,i}|$
are Rayleigh distributed, $i=1,\ldots,N$, with parameter $\sigma_{h,b}^{2}$,
resp. $\sigma_{h,e}^{2}$. We assume that Bob has a good $\mathsf{SNR}$,
but that $\sigma_{b}^{2}=N_{0}<<N_{1}=\sigma_{e}^{2}$, so that Eve
has a poor $\mathsf{SNR}$ with respect to Bob.

The transmitted codeword $\xv\in\RR^{n}$ comes from a lattice $\Lambda_{b}$
intended to Bob, that is 
\[
\xv=M_{b}\uv,~\uv\in\ZZ^{n}
\]
where $M_{b}$ is the generator matrix of the lattice $\Lambda_{b}$.
We can rewrite the channel accordingly: 
\begin{equation}
\begin{array}{ccl}
\yv & = & \diag(\hv_{b})M_{b}\uv+\vv_{b}\\
\zv & = & \diag(\hv_{e})M_{b}\uv+\vv_{e},\end{array}\label{eq:chan}\end{equation}
and we set 
\[
M_{b,\hv_{b}}=\diag(\hv_{b})M_{b},~M_{b,\hv_{e}}=\diag(\hv_{e})M_{b}
\]
which can be interpreted as the generator matrix of the lattice $\Lambda_{b,\hv_{b}}$,
resp. $\Lambda_{b,\hv_{e}}$. In words, these are the lattice intended
to Bob seen through Bob's, resp. Eve's channel.

Coset encoding is used, namely the lattice $\Lambda_{b}$ is partitioned
into a union of disjoint cosets of the form 
\[
\Lambda_{e}+\cv,
\]
with $\Lambda_{e}$ a sublattice of $\Lambda_{b}$ and $\cv$ an
$n$-dimensional vector. To send $k$ bits $\sv$ of data, we need
$2^{k}$ cosets 
\[
\Lambda_{b}=\cup_{j=1}^{2^{k}}(\Lambda_{e}+\cv_{j})
\]
to be labelled by 
\[
\sv\mapsto\Lambda_{e}+\cv_{j(\sv)}.
\]
Alice then randomly chooses a point $\xv\in\Lambda_{e}+\cv_{j(\sv)}$
and sends it over the wiretap channel. This is equivalent to choose
a random vector $\rv\in\Lambda_{e}$. The transmitted lattice point
$\xv\in\Lambda_{b}$ is finally of the form 
\begin{equation}
\xv=\rv+\cv\in\Lambda_{e}+\cv.\label{eq:xsent}
\end{equation}
We can as above set 
\[
M_{e,\hv_{b}}=\diag(\hv_{b})M_{e},~M_{e,\hv_{e}}=\diag(\hv_{e})M_{e}
\]
which are the generator matrix of the lattice $\Lambda_{e,\hv_{b}}$,
resp. $\Lambda_{e,\hv_{e}}$ corresponding to the lattice $\Lambda_{e}$
twisted by the channel of Bob, resp Eve. Bits are transmitted by Alice
at a rate equal to $R=R_{s}+R_{r}$ where $R_{s}$ is the secrecy
rate of this transmission and $R_{r}$ is the rate of random bits.

The parameters involved are: 
\begin{itemize}
\item $\Lambda_{b}$ is the lattice intended for Bob, 
\item $\Lambda_{e}$ is a sublattice of $\Lambda_{b}$ that encodes the
random bits intended for Eve, 
\item $n$ is the dimension of both lattices, 
\item $\mathcal{V}\left(\Lambda_{b}\right)$ (resp. $\mathcal{V}\left(\Lambda_{e}\right)$)
is the fundamental parallelotope of $\Lambda_{b}$ (resp. $\Lambda_{e}$), 
\item $\vol(\Lambda_{b})$ (resp. $\vol(\Lambda_{e})$) is the volume of $\Lambda_{b}$
(resp. $\Lambda_{e}$) where by definition \[
\vol(\Lambda_{b})=\int_{\Vc(\Lambda_{b})}d\xv=\det(M_{b}M_{b}^{T})^{1/2},\]
 
\item the unnormalized second moment $\Uc(\Lambda_{b})$ is 
\[
\Uc(\Lambda_{b})=\int_{\Vc(\Lambda_{b})}||\xv||^{2}d\xv.
\]
\end{itemize}

\subsection{Gaussian channels}

Recall from \cite{sec-gain} that the probability $P_{c,b}$ of Bob's
(resp. $P_{c,e}$ of Eve's) correct decision in doing coset decoding
when $\Lambda_{b}$ is sent over a Gaussian channel is: \begin{eqnarray*}
P_{c,b} & = & \frac{1}{(\sqrt{2\pi}\sigma_{b})^{n}}\sum_{\rv\in\Lambda_{e}}\int_{\Vc(\Lambda_{b}+\rv)}e^{-\Vert\uv\Vert^{2}/2\sigma_{b}^{2}}d\uv\\
P_{c,e} & = & \frac{1}{(\sqrt{2\pi}\sigma_{e})^{n}}\sum_{\rv\in\Lambda_{e}}\int_{\Vc(\Lambda_{b}+\rv)}e^{-\Vert\uv\Vert^{2}/2\sigma_{e}^{2}}d\uv.\end{eqnarray*}

Since $\Lambda_{b}$ is designed for Bob to correctly decode, the
received point is most likely to be in the coset with $\rv={\bf 0}$,
so that \begin{equation}
P_{c,b}\simeq\frac{1}{(\sqrt{2\pi}\sigma_{b})^{n}}\int_{\Vc(\Lambda_{b})}e^{-\Vert\uv\Vert^{2}/2\sigma_{b}^{2}}d\uv.
\label{eq:Pcb}
\end{equation}

As for Eve, $\sigma_{e}$ is assumed larger than $\sigma_{b}$, so
we need to take into account the cosets where $\rv\neq{\bf 0}$, By
writing $\uv=\wv+\rv$, $\wv\in\Lambda_{b}$, a Taylor expansion of 
$e^{-||\wv+\rv||^{2}/2\sigma_{e}^{2}}$ at order 2 gives 
\[
\left(1+\frac{-1}{\sigma_{e}^{2}}\langle\rv,\wv\rangle+\frac{-1}{2\sigma_{e}^{2}}||\wv||^{2}+\frac{1}{2\sigma_{e}^{4}}\langle\rv,\wv\rangle^{2}\right)+O\left(\frac{1}{\sigma_{e}^{4}}\right)
\]
and we get, by neglecting $O\left(\frac{1}{\sigma_{e}^{4}}\right)$, that
\begin{eqnarray*}
&&\sum_{\rv\in\Lambda_{e}}\int_{\Vc(\Lambda_{b}+\rv)}
e^{-||\uv||^{2}/2\sigma_{e}^{2}}d\uv\\
&\simeq&\sum_{\rv\in\Lambda_{e}}e^{-||\rv||^{2}/2\sigma_{e}^{2}}
\left(\vol(\Lambda_{b})-\frac{\Uc(\Lambda_{b})}{2\sigma_{e}^{2}}\right)
\end{eqnarray*}
noticing that 
$\sum_{\rv\in\Lambda_{e}}\int_{\Vc(\Lambda_{b})}\langle\rv,\wv\rangle d\wv=0$
since the sum is over $\rv\in\Lambda_{e}$, and for each $\rv$ in
$\Lambda_{e}$, $-\rv$ is also in $\Lambda_{e}$.

The probability of making a correct decision for Eve is then 
\[
P_{c,e}\simeq\frac{1}{(2\pi\sigma_{e}^2)^{n/2}}
\sum_{\rv\in\Lambda_{e}}e^{-\Vert\rv\Vert^{2}/2\sigma_{e}^{2}}
\left(\vol(\Lambda_{b})-\frac{\Uc(\Lambda_{b})}{2\sigma_{e}^{2}}\right)
\]
and the goal is then to minimize 
\[
\sum_{\rv\in\Lambda_{e}}e^{-\Vert\rv\Vert^{2}/2\sigma_{e}^{2}}.
\]
By further neglecting the terms in $O\left(\frac{1}{\sigma_e^2}\right)$, we further 
simplify Eve's probability of correct decision to
\begin{equation}
P_{c,e}\simeq\frac{\vol(\Lambda_{b})}{(2\pi\sigma_{e}^2)^{n/2}}
\sum_{\rv\in\Lambda_{e}}e^{-\Vert\rv\Vert^{2}/2\sigma_{e}^{2}}.
\label{eq:Pce}\end{equation}

%
%

\section{Code design criterion: Fast Fading Channels}

For a given realization of the fading $\hv$, the channel (\ref{eq:chan})
can be seen as the Gaussian wiretap channel \begin{equation}
\begin{array}{ccl}
\yv & = & M_{b,\hv_{b}}\uv+\vv_{b}\\
\zv & = & M_{b,\hv_{e}}\uv+\vv_{e},
\end{array}\label{eq:chanfad}\end{equation}
and we note that for $\rv\in\Lambda_{e,\hv_{e}}$
\begin{equation}\label{eq:rv}
\Vert\rv\Vert^{2} = \Vert \diag(\hv_e)M_e\uv\Vert^2= \sum_{i=1}^{n}|h_{e,i}x_{i}|^{2},
\end{equation}
with $\uv\in\ZZ^n$ and $\xv\in\Lambda_e$.

Since probability computations for Bob, which is the classical problem of transmitting over a fast Rayleigh fading channel, have been extensively studied in 
the literature (e.g. \cite[sec~2.3]{Oggier-2}), we focus on Eve.

\subsection{Eve's probability of correct decision}

The probability of Eve correctly decoding on channel (\ref{eq:chanfad})
is from (\ref{eq:Pce}), for a given fading realization 
\begin{equation}
P_{c,e,\hv_{e}}\simeq\left(\frac{1}{2\pi\sigma_{e}^{2}}\right)^{n/2}\mbox{Vol}(\Lambda_{b,\hv_{e}})\sum_{\mathbf{r}\in\Lambda_{e,\hv_{e}}}e^{-\frac{\left\Vert \mathbf{r}\right\Vert ^{2}}{2\sigma_{e}^{2}}}.\label{eq:pceh}\end{equation}
As \[
\mbox{Vol}(\Lambda_{b,\hv_{e}})=
\prod_{i=1}^{n}\left|h_{e,i}\right|\mbox{Vol}(\Lambda_{b})
\]
and using (\ref{eq:rv}), we get 
\begin{equation}
\sum_{\mathbf{r}\in\Lambda_{e,\hv_{e}}}e^{-\frac{\left\Vert \mathbf{r}\right\Vert ^{2}}{2\sigma_{e}^{2}}}=\sum_{\xv\in\Lambda_{e}}e^{-\frac{\sum_{i=1}^{n}\left|h_{e,i}x_{i}\right|^{2}}{2\sigma_{e}^{2}}},
\label{eq:theta-he}\end{equation}
yielding the following approximate expression for $P_{c,e,\hv_{e}}$
\begin{align}
 & \left(\frac{1}{2\pi\sigma_{e}^{2}}\right)^{n/2}\mbox{Vol}(\Lambda_{b})\prod_{i=1}^{n}\left|h_{e,i}\right|\sum_{\xv\in\Lambda_{e}}e^{-\frac{\sum_{i=1}^{n}\left|h_{e,i}x_{i}\right|^{2}}{2\sigma_{e}^{2}}}\nonumber \\
= & \left(\frac{1}{2\pi\sigma_{e}^{2}}\right)^{n/2}\mbox{Vol}(\Lambda_{b})\sum_{\xv\in\Lambda_{e}}\prod_{i=1}^{n}\left(\left|h_{e,i}\right|e^{-\frac{\left|h_{e,i}x_{i}\right|^{2}}{2\sigma_{e}^{2}}}\right)\label{eq:pce-he}.\end{align}

The average probability $\bar{P}_{c,e} $ of correct decision is now: 
\begin{eqnarray}
&\!\!&\!\!\!\mathbb{E}_{\hv_{e}}\left[P_{c,e,\hv_{e}}\right]\nonumber \\
&\simeq\!\! & \!\!\!\left(\frac{1}{2\pi\sigma_{e}^{2}}\right)^{n/2}\mathbb{E}_{\hv_{e}}\left[\mbox{Vol}(\Lambda_{b})\sum_{\xv\in\Lambda_{e}}\prod_{i=1}^{n}\left(\left|h_{e,i}\right|e^{-\frac{\left|h_{e,i}x_{i}\right|^{2}}{2\sigma_{e}^{2}}}\right)\right]\nonumber \\
&=\!\! &\!\!\! \left(\frac{1}{2\pi\sigma_{e}^{2}}\right)^{n/2}\mbox{Vol}(\Lambda_{b})\sum_{\xv\in\Lambda_{e}}\mathbb{E}_{\hv_{e}}\left[\prod_{i=1}^{n}\left(\left|h_{e,i}\right|e^{-\frac{\left|h_{e,i}x_{i}\right|^{2}}{2\sigma_{e}^{2}}}\right)\right]\nonumber \\
&= \!\!& \!\!\!\left(\frac{1}{2\pi\sigma_{e}^{2}}\right)^{n/2}\mbox{Vol}(\Lambda_{b})\sum_{\xv\in\Lambda_{e}}\prod_{i=1}^{n}\underbrace{\mathbb{E}_{\hv}\left(|h_{e,i}|e^{-\frac{\left|h_{e,i}x_{i}\right|^{2}}{2\sigma_{e}^{2}}}\right)}_{\mathcal{F}}\label{eq:Pce-with-F}
\end{eqnarray}
since the $|h_{e,i}|$ are independently distributed, $i=1,\ldots,n$.
Set $\rho_{i}=|h_{e,i}|$ which is Rayleigh distributed with parameter
$\sigma_{h,e}^{2}$ and pdf \[
f(\rho_{i},\sigma_{h,e}^{2})=\frac{\rho_{i}}{\sigma_{h,e}^{2}}e^{-\frac{\rho_{i}^{2}}{2\sigma_{h,e}^{2}}}.\]
 Thus \begin{eqnarray*}
\mathcal{F} & = & \frac{1}{\sigma_{h,e}^{2}}\int_{0}^{\infty}\rho_{i}e^{-\frac{\rho_{i}^{2}|x_{i}|^{2}}{2\sigma_{e}^{2}}}\rho_{i}e^{-\frac{\rho_{i}^{2}}{2\sigma_{h,e}^{2}}}d\rho_{i}\\
 & = & \frac{1}{\sigma_{h,e}^{2}}\int_{0}^{\infty}\rho_{i}^{2}e^{-\rho_{i}^{2}\left(\frac{\left|x_{i}\right|^{2}}{2\sigma_{e}^{2}}+\frac{1}{2\sigma_{h,e}^{2}}\right)}d\rho_{i}\\
 & = & \frac{1}{\sigma_{h,e}^{2}}\frac{\sqrt{\pi}}{4\left(\frac{\left|x_{i}\right|^{2}}{2\sigma_{e}^{2}}+\frac{1}{2\sigma_{h,e}^{2}}\right)^{3/2}}\end{eqnarray*}
 since for $a>0$, we have \[
\intop_{0}^{+\infty}x^{2}e^{-ax^{2}}dx=\frac{\sqrt{\pi}}{4a^{\frac{3}{2}}}.\]

Thus $\sum_{\xv\in\Lambda_{e}}\prod_{i=1}^n\mathcal{F}$ in (\ref{eq:Pce-with-F}) becomes 
\begin{eqnarray*}
&&
\sum_{\xv\in\Lambda_{e}}\left(\frac{\sqrt{\pi}}{4\sigma_{h,e}^{2}}\right)^{n}\prod_{i=1}^{n}\frac{1}{\left(\frac{1}{2\sigma_{h,e}^{2}}+\frac{|x_{i}|^{2}}{2\sigma_{e}^{2}}\right)^{\frac{3}{2}}}\\
&=&\sum_{\xv\in\Lambda_{e}}\left(\frac{\sqrt{\pi}}{4\sigma_{h,e}^{2}}\right)^{n}(2\sigma_{h,e}^{2})^{3n/2}\prod_{i=1}^{n}\frac{1}{\left(1+\left|x_{i}\right|^{2}\frac{\sigma_{h,e}^{2}}{\sigma_{e}^{2}}\right)^{\frac{3}{2}}}\\
&=&\sum_{\xv\in\Lambda_{e}}\left(\frac{\sqrt{\pi}\sigma_{h,e}}{\sqrt{2}}\right)^{n}\prod_{i=1}^{n}\frac{1}{\left(1+\left|x_{i}\right|^{2}\frac{\sigma_{h,e}^{2}}{\sigma_{e}^{2}}\right)^{\frac{3}{2}}}\\
\end{eqnarray*}
and (\ref{eq:Pce-with-F}) can be rewritten as
\[
\bar{P}_{c,e}\simeq
\left(\frac{\sigma_{h,e}}{2\sigma_{e}}\right)^{n}\mbox{Vol}(\Lambda_{b})
\sum_{\xv\in\Lambda_{e}}\prod_{i=1}^{n}\frac{1}{\left(1+\left|x_{i}\right|^{2}\frac{\sigma_{h,e}^{2}}{\sigma_{e}^{2}}\right)^{\frac{3}{2}}}.
\]
Now, let $\gamma_{e}$ denote Eve's average $\mathsf{SNR}$ defined
as 
\begin{equation}\label{eq:gammae}
\gamma_{e}=\frac{\sigma_{h,e}^{2}}{\sigma_{e}^{2}}.
\end{equation}
 We finally get \begin{equation}
\boxed{\bar{P}_{c,e}\simeq\left(\frac{\gamma_{e}}{4}\right)^{\frac{n}{2}}\mbox{Vol}(\Lambda_{b})\sum_{\xv\in\Lambda_{e}}\prod_{i=1}^{n}\frac{1}{\left(1+\left|x_{i}\right|^{2}\gamma_{e}\right)^{\frac{3}{2}}}}\label{eq:Pce-final}\end{equation}

As $\bar{P}_{c,e}$ is the average probability of correct decision
for Eve, it has to be minimized. We remark that the terms inside the
summation in (\ref{eq:Pce-final}) are very similar to the terms we
have when we express the error probability on a Rayleigh fast fading
channel \cite{Oggier-2}. We further have \begin{align}
\prod_{i=1}^{n}\frac{1}{\left(1+\gamma_{e}|x_{i}|^{2}\right)^{\frac{3}{2}}} & =\prod_{i=1}^{n}\frac{1}{\gamma_{e}^{3/2}\left(\frac{1}{\gamma_{e}}+|x_{i}|^{2}\right)^{\frac{3}{2}}}\nonumber \\
 & \simeq\frac{1}{\gamma_{e}^{\frac{3}{2}d_{\xv}}}\prod_{i\in\mathcal{J}_{\xv}}\frac{1}{\left|x_{i}\right|^{3}}\label{eq:pairwise-asymp}\end{align}
where $\gamma_{e}$ is big enough to consider $1/\gamma_{e}$ as negligible%
\footnote{This assumption is realistic since $\Lambda_{e}$ is a lattice which
should be {}``perfectly'' decoded by Eve.%
} and $\mathcal{J}_{\xv}$ is the set of indices $i$ such that $x_{i}\neq0$
and $d_{\xv}=\left|\mathcal{J}_{\xv}\right|$ is called the diversity
of $\xv$. We have that $d_{\xv}$ is at most $n$, and if it is $n$
for all $\xv\in\Lambda_{e}$, then we have a full diversity lattice
$\Lambda_{e}$ \[
d_{\xv}=n,\forall\xv\in\Lambda_{e}.\]
 In this case, using (\ref{eq:Pce-final}) and (\ref{eq:pairwise-asymp}),
we derive \begin{eqnarray*}
\bar{P}_{c,e} & \simeq & \left(\frac{\gamma_{e}}{4}\right)^{\frac{n}{2}}\frac{1}{\gamma_{e}^{3n/2}}\mbox{Vol}(\Lambda_{b})\sum_{x\in\Lambda_{e}}\prod_{i=1}^{n}\frac{1}{|x_{i}|^{3}}\\
 & = & \left(\frac{1}{4\gamma_{e}^{2}}\right)^{\frac{n}{2}}\mbox{Vol}(\Lambda_{b})\sum_{x\in\Lambda_{e}}\prod_{i=1}^{n}\frac{1}{|x_{i}|^{3}}.\end{eqnarray*}

\subsection{Full-diversity algebraic lattices}

Full-diversity lattices can be obtained using algebraic lattices \cite{Oggier-2},
that is lattices obtained by embedding the ring of integers of a number
field. Let $K/\QQ$ be a number field of degree $n$ with embeddings
$\sigma_{1},\ldots,\sigma_{n}$ into $\CC$, and denote by $\Oc_{K}$
its ring of integers. We assume that the lattice $\Lambda_{e}$ is
obtained via the canonical embedding of either $\Oc_{K}$ or an integral
ideal $\mathcal{I}$ of $\Oc_{K}$. In that case, $x_{i}=\sigma_{i}(x)$
for $x\in\Oc_{K}$. Then \[
\bar{P}_{c,e}\simeq\left(\frac{1}{4\gamma_{e}^{2}}\right)^{\frac{n}{2}}\mbox{Vol}(\Lambda_{b})\sum_{x\in\Oc_{K}}\frac{1}{|N_{K/\mathbb{Q}}(x)|^{3}}.\]
 
%
%

\section{Code design criterion: Block fading channels}

We now consider the case when the channel between Alice and Bob, resp. 
Eve, is block fading with coherence time $L$, instead of being fast fading, 
that is:
\begin{equation}\label{eq:blockfading}
\begin{array}{ccl}
Y & = & \diag(\hv_{b})X+V_{b}\\
Z & = & \diag(\hv_{e})X+V_{e},
\end{array}
\end{equation}
where the transmitted signal $X$ is a $n\times L$ matrix, $V_{b}$ and $V_{e}$
are $n\times L$ matrices denoting the Gaussian noise at Bob, respectively Eve's side, both with coefficients zero mean, and respective variance $\sigma_{b}^{2}$ and 
$\sigma_{e}^{2}$. The fading matrices are given explicitly in (\ref{eq:fadingmatrix}). 
When $L=1$, we are back to the fast fading case.

In the setting of (\ref{eq:blockfading}), we assume that the fading is constant
over $L$ time slots and that the channel coefficients $h_{b,1}\ldots,h_{b,n}$, 
resp. $h_{e,1},\ldots,h_{e,n}$, on the $n$ parallel paths from Alice to Bob, resp. 
Eve, are supposed independent. In order to focus on the $Ln-$dimensional 
lattice structure of the transmitted signal, we vectorize the received
signal (\ref{eq:blockfading}) and obtain
\begin{eqnarray*}
\mathsf{vec}\left(Y\right)&=&\mathsf{vec}\left(\diag\left(\hv_{b}\right)X\right)+\mathsf{vec}\left(V_{b}\right)\\
&=&
\!\!\!\left(\!\!\!
\begin{array}{ccc}
\diag(\hv_b)&&\\
&\ddots& \\
&& \diag(\hv_b)
\end{array}\!\!\!
\right)\vect(X)+\vect(V_b)\\
\mathsf{vec}\left(Z\right)&=&\mathsf{vec}\left(\diag\left(\hv_{e}\right)X\right)+\mathsf{vec}\left(V_{e}\right)\\
&=&
\!\!\!\left(\!\!\!
\begin{array}{ccc}
\diag(\hv_e)&&\\
&\ddots& \\
&& \diag(\hv_e)
\end{array}\!\!\!
\right)\vect(X)+\vect(V_e).
\end{eqnarray*}
We now interpret the $n\times L$ codeword $X$ as coming from a lattice by writing
\begin{equation}\label{eq:vec}
\vect(X)=M_b\uv,\mbox{ resp. }\vect(X)=M_e\uv
\end{equation}
where $\uv\in\ZZ^{Ln}$ and $M_b$ (resp. $M_e$) denotes the $Ln\times Ln$ generator matrix of the lattice intended to Bob (resp. Eve). Thus in what follows, by a lattice 
point $\xv\in\Lambda_b$ (resp. $\Lambda_e$), we mean that
\[
\xv=\vect(X)
\] 
with $\vect(X)$ as (\ref{eq:vec}).

By setting as for the fast fading case
\begin{eqnarray*}
M_{b,\hv_b}&=&\diag(\diag(\hv_b),\ldots,\diag(\hv_b))M_b,\\
M_{b,\hv_e}&=&\diag(\diag(\hv_e),\ldots,\diag(\hv_e))M_b
\end{eqnarray*}
we can rewrite (\ref{eq:blockfading}) as
\begin{eqnarray*}
\vect(Y) & = & M_{b,\hv_{b}}\uv+\vect(V_{b})\\
\vect(Z) & = & M_{b,\hv_{e}}\uv+\vect(V_{e}),
\end{eqnarray*}
where $M_{b,\hv_{b}}$, resp. $M_{b,\hv_e}$ can be interpreted as the lattice 
generators of the lattices $\Lambda_{b,\hv_{b}}$, resp. $\Lambda_{b,\hv_{e}}$
and thus we get in particular for Eve 
\[
\mbox{Vol}(\Lambda_{b,\hv_{e}})=\left(\prod_{i=1}^{n}\left|h_{e,i}\right|\right)^{L}\mbox{Vol}(\Lambda_{b}).
\]

\subsection{Eve's probability of correct decision}

First, we have from (\ref{eq:Pce}) that
\begin{eqnarray*}
P_{c,e}&\simeq& \left(\frac{1}{2\pi\sigma_{e}^{2}}\right)^{\frac{Ln}{2}}
\vol(\Lambda_{b,\hv_e})\sum_{\rv\in\Lambda_{e,\hv_e}}e^{-||\rv||^2/2\sigma_e^2}\\
&=& 
\frac{\mbox{Vol}(\Lambda_{b})}{(2\pi\sigma_{e}^{2})^{\frac{Ln}{2}}}
\left(\prod_{i=1}^{n}\left|h_{e,i}\right|\right)^{L}
\sum_{\xv\in\Lambda_e}e^{-\sum_{j=1}^{Ln}|h_{e,j}x_j|^2/2\sigma_e^2}
\end{eqnarray*}
where $\Lambda_{e,\hv_e}$ is the lattice with generator matrix 
$M_{e,\hv_e}=\diag(\diag(\hv_e),\ldots,\diag(\hv_e))M_e$, $\xv=\vect(X)$ as 
explained in (\ref{eq:vec}) and $||\rv||^2$ is computed as in (\ref{eq:rv}). 
Since $M_{e,\hv_e}$ contains $L$ copies of $\diag(\hv_e)$, we can further adopt a 
double indexing for coefficients of $\xv$ and write
\[
\sum_{j=1}^{Ln}|h_{e,j}x_j|^2=\sum_{j=1}^{L}\sum_{i=1}^n|h_{e,i}x_{ij}|^2
=\sum_{i=1}^n|h_{e,i}|^2\sum_{j=1}^{L}|x_{ij}|^2.
\]
Note for further usage that since $\xv=\vect(X)$, $x_{ij}$ actually corresponds 
to the $(i,j)$ coefficient of $X$, and $\sum_{j=1}^{L}|x_{ij}|^2$ is a summation over 
the $L$ components of the $i$th row of $X$, that we denote by 
$\xv_i=(x_{i1},\ldots,x_{iL})$. 

The average probability of correct decision for Eve is then
\begin{equation}\label{eq:pce-block}
\bar{P}_{c,e} 
=\frac{\mbox{Vol}(\Lambda_{b})}{(2\pi\sigma_{e}^{2})^{\frac{Ln}{2}}}\nonumber \\
\sum_{\xv\in\Lambda_{e}}\prod_{i=1}^{n}\underbrace{\mathbb{E}_{\hv}\left(|h_{e,i}|^{L}e^{-\frac{\left|h_{e,i}\right|^{2}\sum_{j=1}^{L}\left|x_{ij}\right|^{2}}{2\sigma_{e}^{2}}}\right)}_{\mathcal{F}}
\end{equation}
where 
\begin{eqnarray*}
\mathcal{F} & = & \frac{1}{\sigma_{h,e}^{2}}\int_{0}^{\infty}\rho_{i}^Le^{-\frac{\rho_{i}^{2}\sum_{j=1}^L|x_{ij}|^{2}}{2\sigma_{e}^{2}}}\rho_{i}e^{-\frac{\rho_{i}^{2}}{2\sigma_{h,e}^{2}}}d\rho_{i}\\
 & = & \frac{1}{\sigma_{h,e}^{2}}\int_{0}^{\infty}\rho_{i}^{L+1}e^{-\rho_{i}^{2}\left(\frac{\sum_{j=1}^L\left|x_{ij}\right|^{2}}{2\sigma_{e}^{2}}+\frac{1}{2\sigma_{h,e}^{2}}\right)}d\rho_{i}\\
 & = & 
\frac{1}{\sigma_{h,e}^{2}}\frac{\Gamma\left(\frac{L}{2}+1\right)}{\left(\frac{\left\Vert \mathbf{x}_{i}\right\Vert ^{2}}{2\sigma_{e}^{2}}+\frac{1}{2\sigma_{h,e}^{2}}\right)^{\frac{L}{2}+1}}
\end{eqnarray*}
since for $a>0$, we have 
\[
\int_{0}^{+\infty}x^{L+1}e^{-ax^{2}}dx=\frac{\Gamma\left(\frac{L}{2}+1\right)}{2a^{\frac{L}{2}+1}}.
\]
Now $\sum_{\xv\in\Lambda_{e}}\prod_{i=1}^n\mathcal{F}$ is given, as done earlier, by
\begin{eqnarray*}
\!\!\!\!\!\!&&\!\!\!\!\!\!
\sum_{\xv\in\Lambda_{e}}\left(\frac{\Gamma\left(\frac{L}{2}+1\right)}{2\sigma_{h,e}^{2}}\right)^{n}\prod_{i=1}^{n}\frac{1}{\left(\frac{1}{2\sigma_{h,e}^{2}}+\frac{||\xv_{i}||^{2}}{2\sigma_{e}^{2}}\right)^{\frac{L}{2}+1}}\\
\!\!\!\!\!\!&=&\!\!\!\!\!\!
\sum_{\xv\in\Lambda_{e}}\left(\Gamma\left(\frac{L}{2}+1\right)(2\sigma_{h,e}^{2})^{\frac{L}{2}}
\right)^{n}\prod_{i=1}^{n}\frac{1}{\left(1+||\xv_{i}||^{2}\frac{\sigma_{h,e}^{2}}{\sigma_{e}^{2}}\right)^{\frac{L}{2}+1}}
\end{eqnarray*}
Recall from (\ref{eq:gammae}) that Eve's average $\snr$ is
\[
\gamma_{e}=\frac{\sigma_{h,e}^{2}}{\sigma_{e}^{2}}.
\]
We finally conclude that
\[
\bar{P}_{c,e}\simeq\frac{\gamma_{e}^{\frac{Ln}{2}}\Gamma\left(\frac{L}{2}+1\right)^n}{(2\pi)^{\frac{Ln}{2}}}\mbox{Vol}(\Lambda_{b})\sum_{\xv\in\Lambda_{e}}\prod_{i=1}^{n}\frac{1}{\left(1+\left\Vert \mathbf{x}_{i}\right\Vert ^{2}\gamma_{e}\right)^{\frac{L}{2}+1}}.\]
In the same way as in (\ref{eq:pairwise-asymp}), we can express the
term inside the summation by assuming that Eve's SNR $\gamma_{e}$
is high compared to the minimum distance of $\Lambda_{e}$ and get
\[
\prod_{i=1}^{n}\frac{1}{\left(1+\left\Vert \mathbf{x}_{i}\right\Vert ^{2}\gamma_{e}\right)^{\frac{L}{2}+1}}\simeq\frac{1}{\gamma_{e}^{n\left(\frac{L}{2}+1\right)}}\prod_{i=1}^{n}\frac{1}{\left\Vert \mathbf{x}_{i}\right\Vert ^{L+2}}\]
if we assume that none of the $\left\Vert \mathbf{x}_{i}\right\Vert $
are equal to zero. This corresponds to the case where the $Ln$-dimensional lattice 
$\Lambda_{e}$ has diversity order at least $L(n-1)+1$. Indeed, a diversity of $L(n-1)$ or less means that at most $L(n-1)$ coefficients of a non-zero lattice vector are 
non-zero, thus there could be $L$ zero coefficients, which, if all located on the same 
row $i$, would make $||\xv_i||=0$. This cannot happen if the diversity is at least 
$L(n-1)+1$.

In this case, we derive that
\begin{equation}
\bar{P}_{c,e}\simeq
\left(
\frac{\Gamma\left(\frac{L}{2}+1\right)}{(2\pi)^{\frac{L}{2}}\gamma_{e}}\right)^n
\mbox{Vol}(\Lambda_{b})\sum_{\xv\in\Lambda_{e}}\prod_{i=1}^{n}\frac{1}{\left\Vert \mathbf{x}_{i}\right\Vert ^{L+2}}.\label{eq:pce-block-final}
\end{equation}

\subsection{Full-diversity algebraic lattices}

Again, to make sure that full diversity is achieved, we propose to use 
algebraic lattices. But this time, we need to control the terms in (\ref{eq:pce-block-final}), that is essentially the sum 
\begin{equation}
\sum_{\xv\in\Lambda_{e}}\prod_{i=1}^{n}\frac{1}{\left\Vert \mathbf{x}_{i}\right\Vert ^{L+2}}.
\label{eq:sum-block}
\end{equation}
Let $K/\QQ$ be a number field of degree $n$, with $n$ embeddings $\left(\sigma_{1},\sigma_{2},\ldots,\sigma_{n}\right)$ into $\CC$, and ring of integers 
$\mathcal{O}_{\mathbb{K}}$. Recall that a vector point $\xv\in\Lambda_e$ is obtained 
from $\xv=\vect(X)$, and $X$ is the codeword sent. Let $\xv_1$ be the first row of 
$X$, and take $\mathbf{x}_{i}=\sigma_{i}\left(\mathbf{x}_1\right)$, so that each row 
of $X$ is obtained by conjugating its first row. Alternatively, each column can be 
seen as a lattice point from the algebraic lattice build over $\Oc_k$. In this case, 
it is enough for this lattice to be of diversity $n$ to guarantee that 
$||\xv_i||\neq 0$ for all $i$. Indeed, for every non-zero coefficient of the first row 
$\xv_1$, all the corresponding columns will have non-zero coefficients. Conversely, 
each zero coefficient on the first row gives a column of zeros, and to have  
$||\xv_i||=0$ for one $i$ means to have $||\xv_i||=0$ for all $i$, that is sending 
$X$ containing only zeros. 
Now
\[
\left\Vert \mathbf{x}_{i}\right\Vert^2 
=
\left\Vert \sigma_i(\xv_1)\right\Vert ^2
=
\sum_{j=1}^L \sigma_i(x_{1j})^2\\
=
\sigma_i(\sum_{j=1}^L x_{1j}^2)=
\sigma_i(||\xv_1||^2)
\]
and
\[
\prod_{i=1}^n\left\Vert \mathbf{x}_{i}\right\Vert^2 =
\prod_{i=1}^n\left\Vert \sigma_i(\xv_1)\right\Vert ^2=
\prod_{i=1}^n \sigma_i(||\xv_1||^2)=
N_{K/\QQ}(||\xv_1||^2).
\]
The sum in (\ref{eq:sum-block}) finally becomes
\[
\sum_{\xv\in\Lambda_{e}}\prod_{i=1}^{n}\frac{1}{\left\Vert \sigma_{i}\left(\mathbf{x}_1\right)\right\Vert ^{L+2}}=\sum_{\xv\in\Lambda_{e}}\frac{1}{N_{K/\mathbb{Q}}\left(\left\Vert \mathbf{x}_1\right\Vert ^{2}\right)^{\frac{L}{2}+1}}.
\]

%
%
%
\section{Future Work}

Current and future work naturally involves (i) the analysis of the wiretap MIMO Channel so as to determine the corresponding code design criterion, and (ii) the 
construction of lattices optimized for fast fading wiretap channel, block fading wiretap channel, and finally MIMO wiretap channel.

%
%
\section*{Acknowledgements}

Part of this work was done while J.-C. Belfiore was visiting the Nanyang 
Technological University, Singapore. 
The research of F. Oggier is supported in part by the Singapore National
Research Foundation under Research Grant NRF-RF2009-07 and 
NRF-CRP2-2007-03, and in part by the Nanyang Technological University under Research 
Grant M58110049 and M58110070. 

%
%


\begin{thebibliography}{11}

\bibitem{sec-gain} J.-C. Belfiore and F.~Oggier, {}``Secrecy gain:
a wiretap lattice code design,'' ISITA 2010, 2010. {[}Online{]}.
Available: \url{arXiv:1004.4075v1 [cs.IT]}

\bibitem{unimodular} J.-C. Belfiore and P.~Solé, {}``Unimodular
lattices for the Gaussian Wiretap Channel,'' ITW 2010, Dublin. {[}Online{]}.
Available: \url{arXiv:1007.0449v1 [cs.IT]}

\bibitem{HY-09} X.~He and A.~Yener, {}``Providing secrecy with
structured codes: Tools and applications to two-user {G}aussian
channels,'' July 2009. {[}Online{]}. Available: \url{http://arxiv.org/pdf/0907.5388}

\bibitem{KHMBK-09} D.~Klinc, J.~Ha, S.~McLaughlin, J.~Barros,
and B.~Kwak, {}``{LDPC} codes for the {G}aussian wiretap channel,''
in \emph{Proc. Information Theory Workshop}, October 2009.

\bibitem{Oggier-2} F.~Oggier and E.~Viterbo, {}``Algebraic number
theory and code design for {R}ayleigh fading channels,'' in \emph{Foundations
and Trends in Communications and Information Theory}, 2004, vol.~1,
no.~3, pp. 333--415.

\bibitem{Shlomo} Y.~Liang, H.~Vincent Poor, and S.~Shamai (Shitz),{}
``Information Theoretic Security," in \emph{Foundations 
and Trends in Communications and Information Theory}, 2010, vol.~5, no.~4-5.

\bibitem{Wyner-II} L.~H. Ozarow and A.~D. Wyner, {}``Wire-tap
channel {II},'' \emph{Bell Syst. Tech. Journal}, vol.~63, no.~10,
pp. 2135--2157, December 1984.

\bibitem{TDCMM-07} A.~Thangaraj, S.~Dihidar, A.~R.~Calderbank,
S.~McLaughlin, and J.-M. Merolla, {}``Applications of {LDPC} codes
to the wiretap channel,'' \emph{{IEEE} Trans. Inf. Theory}, vol.~53,
no.~8, August 2007.

\bibitem{Wyner-I} A.~Wyner, {}``The wire-tap channel,'' \emph{Bell.
Syst. Tech. Journal}, vol.~54, October 1975.

\end{thebibliography}
\end{document}